\documentclass[preprint,showpacs,nofootinbib]{revtex4}
\usepackage{graphicx}
\usepackage{amsmath}
\usepackage{hyperref}
\newcommand{\avgb}{\langle B \rangle}
\newcommand{\avgnpart}{\langle N_{\rm part} \rangle}

\begin{document}

\title{Baryon and electric charge stoppings in nuclear collisions and
  the role of strangeness} 
\author{Mason Alexander Ross}
\author{Zi-Wei Lin}
\affiliation{Department of Physics, East Carolina University,
  C-209 Howell Science Complex, Greenville, NC 27858, USA}
\email[]{linz@ecu.edu}
\date{\today}

\begin{abstract}
It has been challenging to quantitatively understand the stopping of
incoming nucleons in nuclear collisions, and recently it has been proposed
that comparing the baryon stopping with electric charge stopping can
help address the question. Here we focus on the 
$B/Q\times Z/A$ ratio, which can strongly depend on rapidity
although its value is one for the full phase space.  We find
that this ratio is very sensitive to the difference between strange
and anti-strange rapidity distributions (the $s-\bar s$ asymmetry),
and slightly more anti-strange quarks at mid-rapidity would lead to a
ratio well below one. This is the case for Zr+Zr and Ru+Ru isobar
collisions at $200A$ GeV from a multi-phase transport (AMPT) model. 
Without the $s-\bar s$ asymmetry, the AMPT model would give a
mid-rapidity $B/Q\times Z/A$ ratio at or above one.  In addition, the
AMPT model gives $B/\Delta Q\times \Delta Z/A<1$ at mid-rapidity for
isobar collisions at all centralities, which strongly contradicts the
recent data from the STAR Collaboration.  We further find that the $B/\Delta
Q\times \Delta Z/A$ ratio is very sensitive to the net-light quark
($u,d$) stoppings, but it is less sensitive to the $s-\bar s$ asymmetry
than the $B/Q\times Z/A$ ratio by a factor of 3.
\end{abstract}
\maketitle

\section{Introduction}

In high energy heavy-ion collisions, nucleons from the incoming nuclei
deposit their energy to form a hot and dense phase of matter
called the quark-gluon plasma (QGP)~\cite{Mishustin:2001ib}.  
As a result, how much the initial nucleons stop or lose energy is
important as it determines the initial energy density and net-baryon
density that will affect the equation of state and the evolution of
the created matter. Although the net baryon number per event is a
conserved quantity, how it is redistributed in the phase space 
including its rapidity
distribution~\cite{Busza:1983rj,NA49:1998gaz,Mishustin:2001ib} is not
well understood. In the traditional QCD picture or the standard model,
each quark carries a baryon number of one-third while each antiquark  
carries negative one-third, while a baryon consists of three valence
quarks that are the carriers of both the net-baryon number and
net-electric charge. On the other hand, the baryon junction model has
been  proposed~\cite{Artru:1974zn,Rossi:1977cy,Kharzeev:1996sq}, where
a baryon is composed of three quarks connected by a Y-shaped 
string junction. Naively, the baryon junction is expected to be more
easily stopped since it does not need to carry a large fraction of
the incoming nucleon's momentum.
 
Recently the ratio $B/Q\times Z/A$ has been proposed~\cite{Xu:2022} as
an observable for studying the baryon stopping mechanism, where $B$
and $Q$ respectively represent the net-baryon and  
net-electric charge  within a given acceptance (e.g., at
mid-rapidity), and $Z$ and $A$ are respectively  the atomic number and 
mass number of the incoming nucleus in symmetric A+A collisions.   
The naive expectation is $B/Q\times Z/A=1$, because this is true for
the full phase space due to the corresponding conservation laws.  
By comparing the experimental data with models that incorporate
different stopping mechanisms, we expect to learn more about the
physics responsible for baryon (and electric charge)
stopping~\cite{Lewis:2022arg,Lv:2023fxk,Lin:2024}.  
As the net-electric charge is very difficult to measure
experimentally, the Zr+Zr and Ru+Ru isobar collisions at RHIC provide
a good opportunity because of the large statistics and cancellation of
certain systematic errors.  
The STAR Collaboration has measured the difference of the net-electric 
charge between Ru+Ru and Zr+Zr isobar collisions ($\Delta Q$) as well
as the ratio $B/\Delta Q\times \Delta
Z/A$~\cite{Ma:2024,STAR:2024lvy}. Again, the naive expectation is  
$B/\Delta Q\times \Delta Z/A=1$ since this is true for the full phase
space.   

Our study here addresses the above ratios between the net-baryon
stopping and net-electric charge stopping. We first examine these
ratios at the parton level. We then present results from A Multi-Phase
Transport (AMPT) Model~\cite{Lin:2004en,Lin:2021mdn}, including the
ratios versus rapidity and centrality at different evolution stages of
the created matter. Recent studies from transport models such as UrQMD 
~\cite{Bass:1998ca,Lewis:2022arg,Lv:2023fxk} and
AMPT~\cite{Lin:2004en,Lewis:2022arg,Lin:2024} have shown
$B/Q\times Z/A < 1$ at  mid-rapidity, meaning more charge
stopping than baryon stopping. 
We shall show that the main cause of this is due to slightly
more anti-strange than strange quarks at mid-rapidity. 
In addition, we investigate the $B/\Delta Q\times \Delta Z/A$ ratio
and its sensitivity to the stopping of different quark flavors.

\section{Baryon and charge stoppings at the parton level}

We now consider the parton phase of the dense matter under the
traditional picture where the baryon and electric charge are carried
by quarks and antiquarks. Let us use $f_i$ as the short notation for
the rapidity distribution $dN/dy$ of quark flavor $i$ after the
collision of two relativistic heavy ions, we then have   
\begin{eqnarray}
f_B &\equiv& \frac{dN_B}{dy} =(f_u-f_{\bar u}+f_d-f_{\bar d}+f_s-f_{\bar s})/3, 
\nonumber \\
f_Q &\equiv& \frac{dN_Q}{dy} =(2f_u-2f_{\bar
    u}-f_d+f_{\bar d} -f_s+f_{\bar s})/3. 
\label{bq}
\end{eqnarray}
For high energy symmetric A+A collisions at a given
centrality, a fraction $p$ of the nucleons in each incoming nucleus
will be participant nucleons and converted into partons.  
Since we neglect the effect of neutron skin in this study, 
the same fraction of protons in each incoming nucleus will interact. 
For each quark flavor, we write its total number as $N_i=\int f_i
\;dy$, and we have $N_s=N_{\bar s}$. When we integrate over the full
phase space, we get
\begin{eqnarray}
&B&=2pA=(N_u-N_{\bar u}+N_d-N_{\bar d})/3, \nonumber \\
&Q&=2pZ=(2N_u-2N_{\bar u}-N_d+N_{\bar d})/3, \nonumber \\
&\rightarrow& B/Q\times Z/A=1. 
\end{eqnarray}
The last relation above may be called the naive expectation, since it
is only valid in full phase space. 
At a given rapidity, $B/Q\times Z/A=f_B/f_Q\times Z/A$ 
depends on several variables in Eq.\eqref{bq}, because 
in general the quark rapidity distributions can all be different,
e.g., $f_u \neq f_d$ and/or $f_s \neq f_{\bar   s}$. 

A useful limit is the isospin-symmetric case where $A=2Z$ (like 
$^{40}\rm Ca$ or $^{20}\rm Ne$), where we expect 
$f_u=f_d \equiv f_q$ and  $f_{\bar u}=f_{\bar d} \equiv f_{\bar q}$,
Eq.\eqref{bq} then reduces to  
\begin{eqnarray}
B \equiv f_B=( 2f_q-2f_{\bar q}+f_s-f_{\bar s} )/3, ~~Q \equiv
  f_Q=(f_q-f_{\bar q}-f_s+f_{\bar s})/3.  
\label{bqSym}
\end{eqnarray}
If $f_s=f_{\bar s}$ (i.e., strange and anti-strange quarks have
the same rapidity distribution), we then have
\begin{eqnarray}
B/Q\times Z/A=1  {\rm~at~any~}y,
\label{bqSym2}
\end{eqnarray}
regardless of the stoppings of net-light quarks (meaning $u,d$ in this
study). Equivalently, if we find $B/Q\times Z/A \neq 1$ at a  certain
rapidity  for symmetric A+A collisions with $A=2Z$, this 
breaking of the naive expectation is caused by the strange 
and anti-strange asymmetry according to Eq.\eqref{bqSym}, where
strange quarks and anti-strange quarks must have different shapes in
the rapidity distribution. We note that the neutron skin 
effect~\cite{Hammelmann:2019vwd,Xu:2021qjw,Ding:2024xxu,Pihan:2024lxw}
will modify some of the above relations.  Nevertheless, the above analysis
clearly demonstrates the essential role that strange quarks play in
the $B/Q\times Z/A$ ratio.
 
\section{Results from a multi-phase transort model}

We now study the stopping of baryon and electric charges with the AMPT
model~\cite{Lin:2004en}, which  allows us to study the 
stoppings at different stage of the dense matter's evolution.  
The string melting AMPT model (AMPT-SM) consists of four main
components: the initial conditions, partonic interactions, conversion
from partonic to hadronic matter via quark coalescence, and hadronic
interactions.   
Note that this version of the model includes the improvement that
conserves the net electric charge of each event
exactly~\cite{Lin:2021mdn}.  
It also includes a quark coalescence with the reversed coalescence
order, where  it searches for baryon or anti-baryon partners before
searching for meson partners. This has been found to lead to better
descriptions of baryons and anti-baryons similar to the new
quark coalescence model~\cite{He:2017tla}, where the order is removed
and a quark has the freedom to form either a meson or a baryon
depending on the coalescence distance. 
Furthermore, the HIJING1 model~\cite{Gyulassy:1994ew} for the AMPT
initial condition orders the $A$ nucleons according to the
z-coordinate values and assigns the first $Z$ nucleons as
protons. Since this artifact affects the initial electric charge
stopping, we have removed the artifact in the AMPT model used for
this study by breaking the artificial association of the z-coordinate
with the nucleon charge. We then apply this AMPT model to isobar 
($A=96$) collisions at $\sqrt {s_{NN}}=200$ GeV. 
Note that in this study we use the Woods-Saxon density
distribution to sample nucleons in the isobar nuclei.  
Including the neutron skin~\cite{Xu:2021qjw} and deformations 
of the isobar nuclei~\cite{Pihan:2024lxw}, which 
is left for future work, could lead to modifications of the electric
charge stopping and consequently the baryon-to-charge ratios. 

\subsection{Baryon and charge stoppings for isobar systems}

\begin{figure}[htb]
\includegraphics[width=0.5\linewidth]{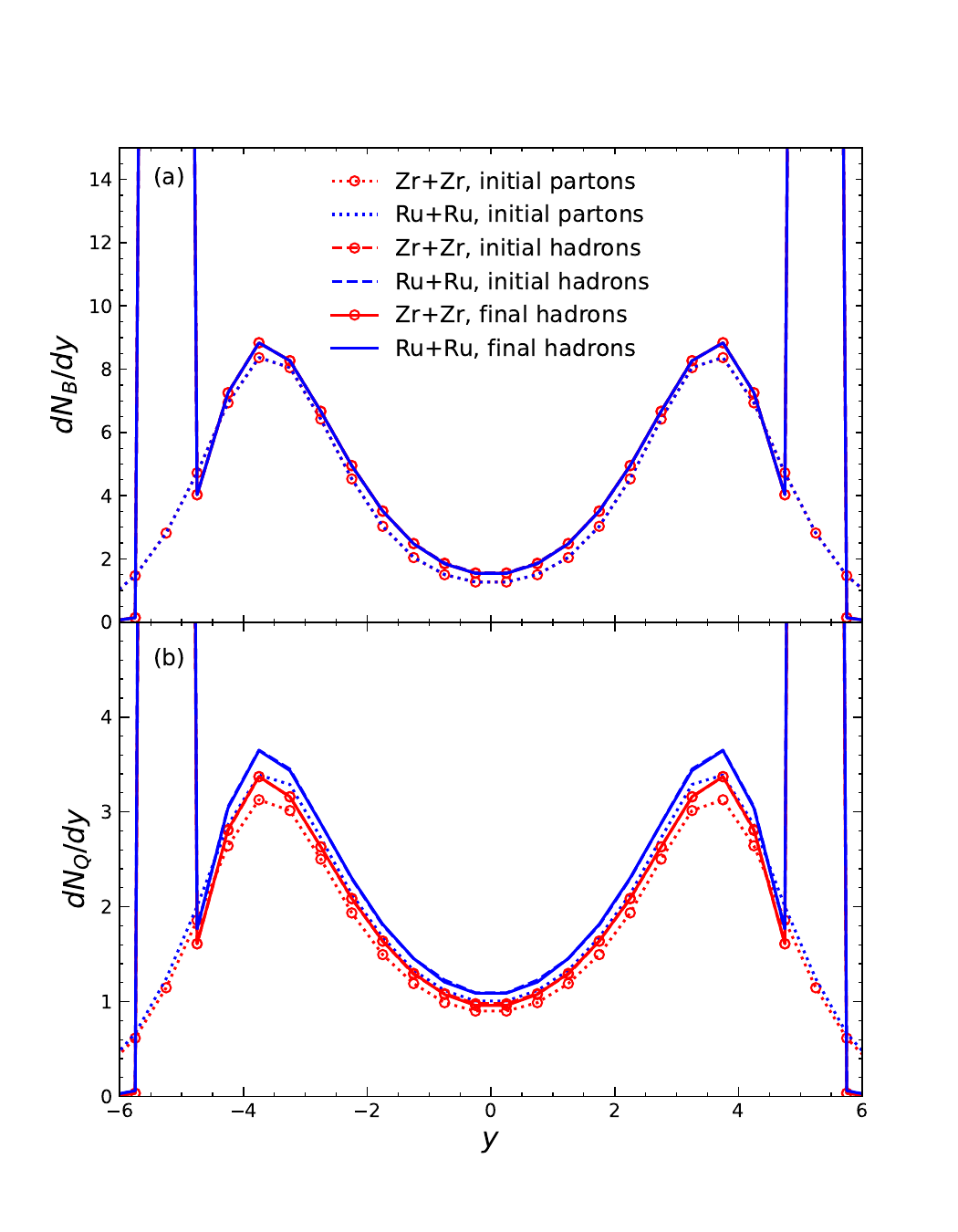}
\caption{The rapidity distributions of (a) the net-baryon number and
  (b) the net-electric charge in minimum bias isobar collisions at
  $\sqrt{s_{\rm NN}}=200A$ GeV from the AMPT-SM model
from initial partons (dotted), initial hadrons (dashed) and final
state hadrons (solid).}
\label{fig1}
\end{figure}

Figure~\ref{fig1} shows the rapidity distributions of the net-baryon
number in panel (a) and the net-electric charge in panel (b) 
from the AMPT-SM model for minimum bias Zr+Zr and Ru+Ru isobar
collisions at $200A$ GeV. 
Solid curves are the results for final state hadrons (i.e., after the
hadron cascade and resonance decays), while dashed curves 
represent the results of initial hadrons (without the hadron cascade
but with resonance decays).  We see that the hadron cascade has
little effect on  these distributions. 
The results at the parton level right after string melting but before
the parton cascade, as calculated via Eq.\eqref{bq}, are also shown as
dotted curves; they are quite similar to the final hadron
distributions (except for the beam rapidity regions). In addition, we
see the expected features that the two isobar systems have essentially
the same net-baryon distributions but different net-electric charge
distributions.

\begin{figure}[htb]
\includegraphics[width=0.5\linewidth]{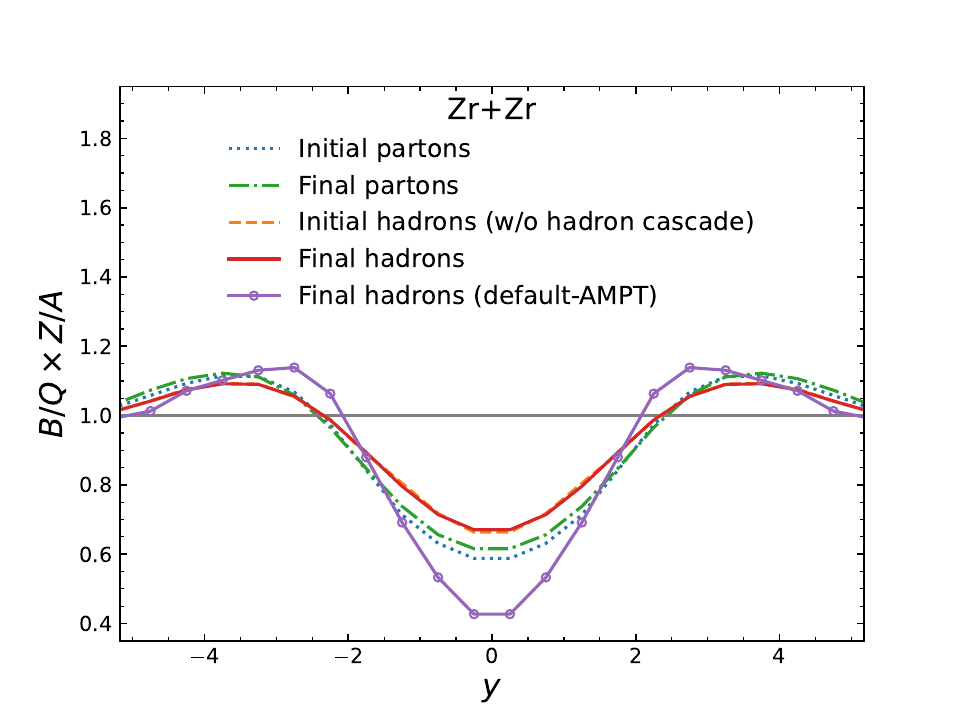}
\caption{The $B/Q \times  Z/A$ ratio from the AMPT-SM model
as a function of rapidity at different stages of the evolution of
minimum bias Zr+Zr collisions; the ratio from the default AMPT model
(curve with circles) is also shown.}
\label{fig2}
\end{figure}

We show in Fig.~\ref{fig2} the $B/Q\times Z/A$ ratio versus 
rapidity at four different stages of the time evolution of minimum
bias Zr+Zr collisions at $200A$ GeV from the AMPT-SM model. 
We see that the rapidity distributions have a similar shape and
are all well below unity at mid-rapidity. The parton cascade has a
small effect on the $B/Q\times Z/A$ rapidity distribution, while the hadron
cascade almost has no effect. 
The result from the default version of the AMPT
model~\cite{Lin:2004en} is also well below unity at mid-rapidity,  
although it has a weaker parton cascade phase but a
stronger hadron cascade phase and uses the Lund 
fragmentation for hadronization instead of quark coalescence. 

\subsection{The role of strangeness on the $B/Q\times Z/A$ ratio}

\begin{figure}[htb]
\includegraphics[width=0.8\linewidth]{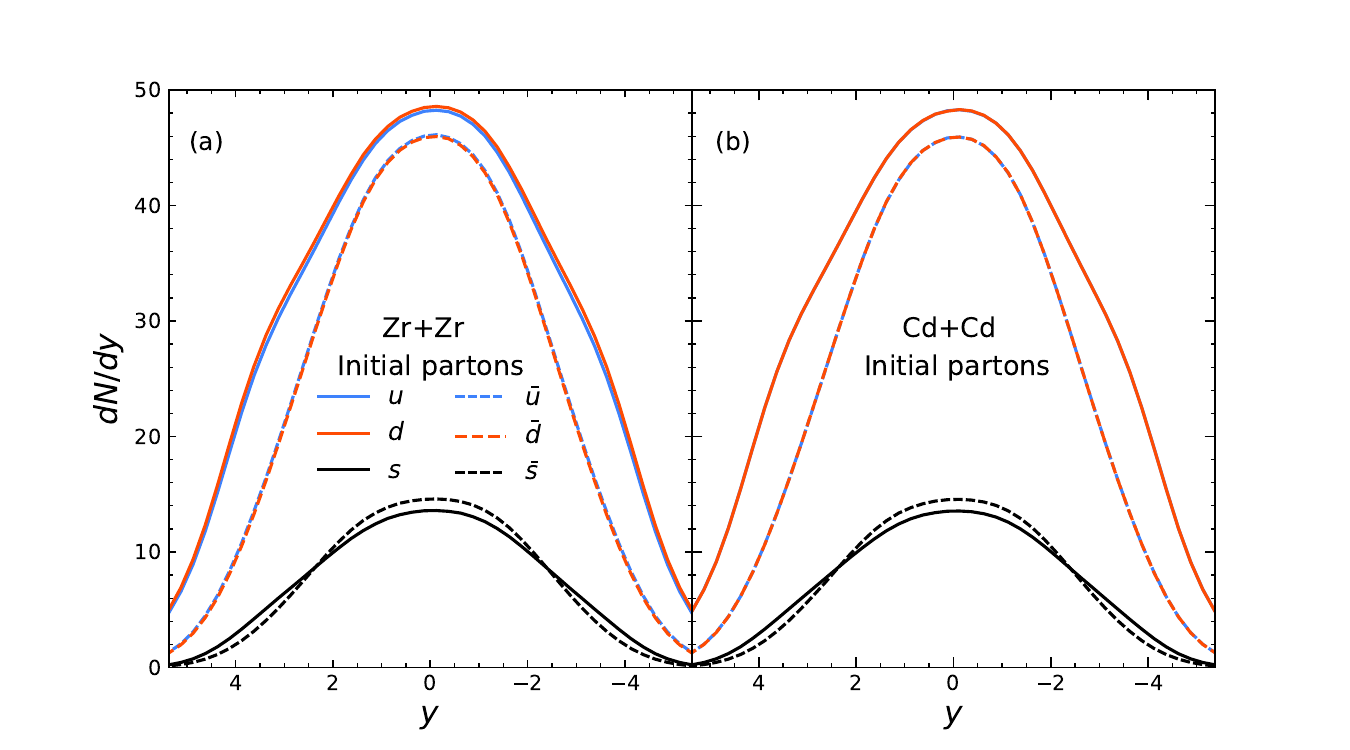}
\caption{Rapidity distributions of initial quarks from the AMPT-SM
  model for minimum bias (a) Zr+Zr and (b) Cd+Cd isobar collisions
at $200A$ GeV.}
\label{fig3}
\end{figure}

To further understand why the naive expectation is badly broken, we
plot in Fig.~\ref{fig3} the rapidity distributions of initial quarks
in Zr+Zr collisions and the hypothetical Cd+Cd isobar 
collisions (at $A=96$), where initial quarks refer to the quarks and
antiquarks produced by the string melting mechanism~\cite{Lin:2004en}
before the parton cascade. Note that the isobar Cd+Cd collisions are
hypothetical since there is no known $^{96}$Cd isotope; we simulate
their collisions to demonstrate  the roles of strangeness and isospin
since the $^{96}$Cd nucleus is isospin-symmetric. 
We see in Fig.~\ref{fig3} a small but obvious difference between
strange and anti-strange quark rapidity distributions, which is
similar for the two collision systems.  Around mid-rapidity ($|y|
\lesssim 2$) there are more anti-strange than strange quarks while the
opposite happens at large rapidities. Around mid-rapidity, this
asymmetry  leads to a negative contribution to the net-baryon number
but a positive contribution to the net electric charge as shown in 
Eq.\eqref{bq}; thus this $s-\bar s$ asymmetry suppresses the
$B/Q\times Z/A$ ratio. 
For the isospin-symmetric $^{96}$Cd system, we see in
Fig.~\ref{fig3}(b) that $u$ and $d$ quarks have the same rapidity
distribution and so do $\bar u$ and $\bar d$ quarks. Therefore, the
difference in the $s$ and $\bar s$ distributions, i.e., $f_s \neq 
f_{\bar s}$, is the only reason for $B/Q\times Z/A \neq 1$ for
Cd+Cd collisions according to Eqs.~\eqref{bqSym}-\eqref{bqSym2}. 

To demonstrate how a seemingly small $s-\bar s$ asymmetry can give a
large deviation of $B/Q\times Z/A$ from unity, let us 
rewrite the quark level Eq.\eqref{bq} as 
\begin{eqnarray}  
B=(f_{net-u}+f_{net-d}+f_{net-s})/3, ~~Q=(2f_{net-u}-f_{net-d}-f_{net-s})/3,  
\end{eqnarray}  
where $f_{net-u} \equiv f_u-f_{\bar u}$ reflects the net-u quark
stopping. For collisions of most nuclei where $Z \sim A/2$, we then get 
\begin{eqnarray} 
B/Q\times Z/A &\simeq& \frac{Z}{A} \left
                 (\frac{f_{net-u}+f_{net-d}}{2f_{net-u}-f_{net-d}}
                 \right   ) \left (1+\frac{3f_{net-s}}{2f_{net-q}} \right )
\label{bq} \\
&{\rm with~}& f_{net-q} \equiv (f_{net-u}+f_{net-d})/2 
\end{eqnarray}
at the first order of $f_{net-s}/f_{net-q}$. 
In particular, for the isospin symmetric $^{96}$Cd+$^{96}$Cd isobar
collisions,  we have $f_{net-u}=f_{net-d}$ and thus get
\begin{eqnarray}
B/Q\times Z/A \simeq 1+\frac{ 3}{2}\frac{f_{net-s}}{f_{net-q}}.
\end{eqnarray}
Although the $s-\bar s$ asymmetry (i.e., $f_{net-s}/f_s$) is quite
small ($\lesssim 10\%$) as shown in Fig.~\ref{fig3},
$f_{net-s}/f_{net-q}$ is much higher than that, and it causes a big
deviation of the mid-rapidity $B/Q\times Z/A$ from unity. 

\begin{figure}[htb]
\includegraphics[width=0.8\linewidth]{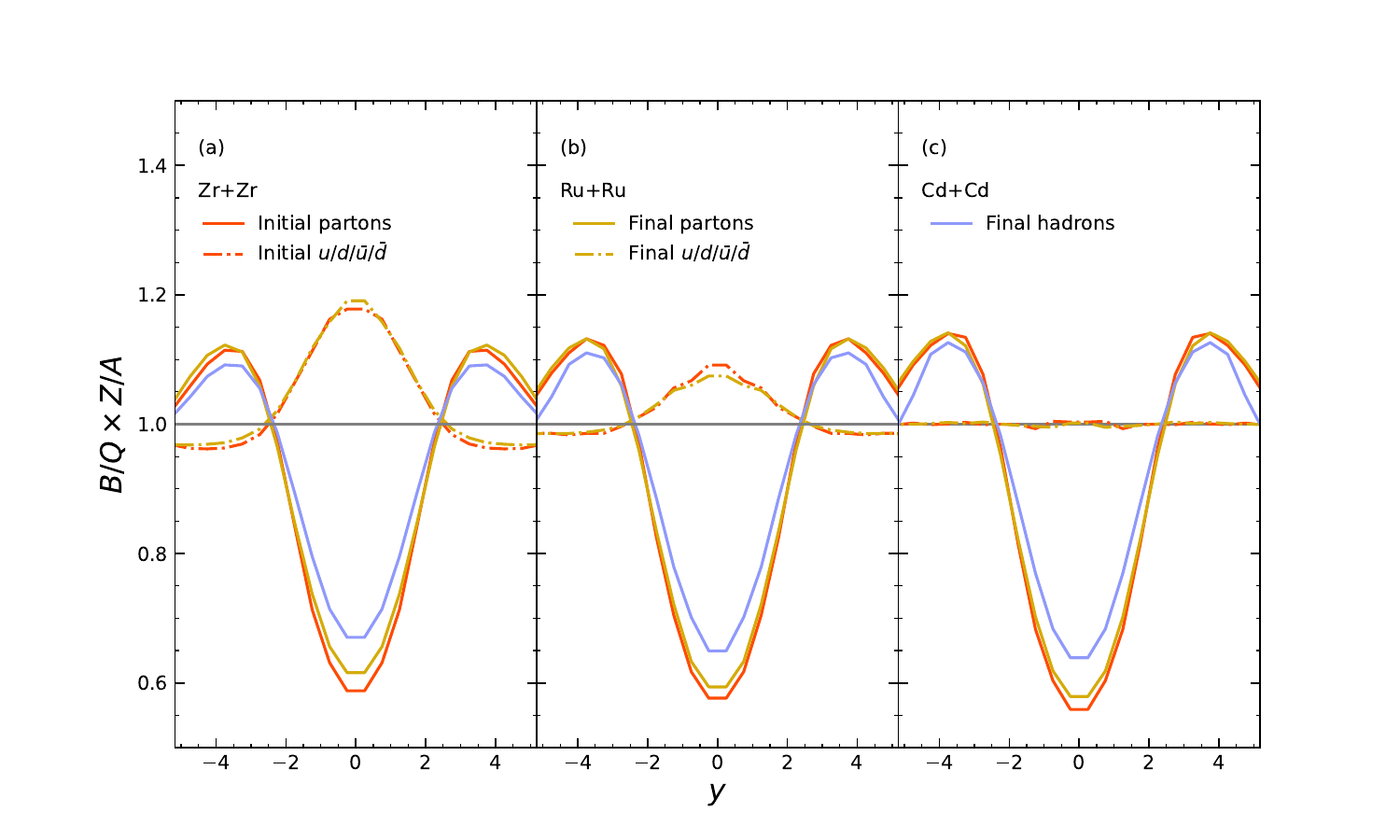}
\caption{The $B/Q\times Z/A$ ratio as a function of rapidity for
  minimum bias collisions of three isobar systems at
  $200A$ GeV from the AMPT-SM model, including results for initial
  partons and final partons with and without the inclusion of strange quarks as well as final state hadrons.}
\label{fig4}
\end{figure}

Figure~\ref{fig4} shows the $B/Q\times Z/A$ ratio as a function of
rapidity for three isobar systems from the AMPT-SM model, 
where the results for initial partons (before the parton cascade), 
final partons (after the parton cascade), and final state hadrons
are shown. When we only include light ($u,d, \bar u, \bar d$)
quarks,  we have $B/Q\times Z/A = 1.0$ at any rapidity
for Cd+Cd collisions, consistent with Eq.\eqref{bqSym2}; 
however,  the $B/Q\times Z/A$ ratio changes drastically and becomes
well below unity around mid-rapidity when we also include
(anti)strange quarks.  From Cd+Cd to Ru+Ru to Zr+Zr collisions, the
isospin asymmetry of the incoming nuclei is increasing bigger, and
consequently the $B/Q\times Z/A$ ratio for initial light quarks
increasingly deviates from unity.  The ratios for all three collision
systems are very close to each 
other, either for initial partons, final partons, or final hadrons, as
long as strange particles are included; therefore, the $s-\bar s$
asymmetry in rapidity is the primary reason for $B/Q\times Z/A<1$ at
mid-rapidity in the AMPT-SM model.

\begin{figure}[htb]
\includegraphics[width=0.8\linewidth]{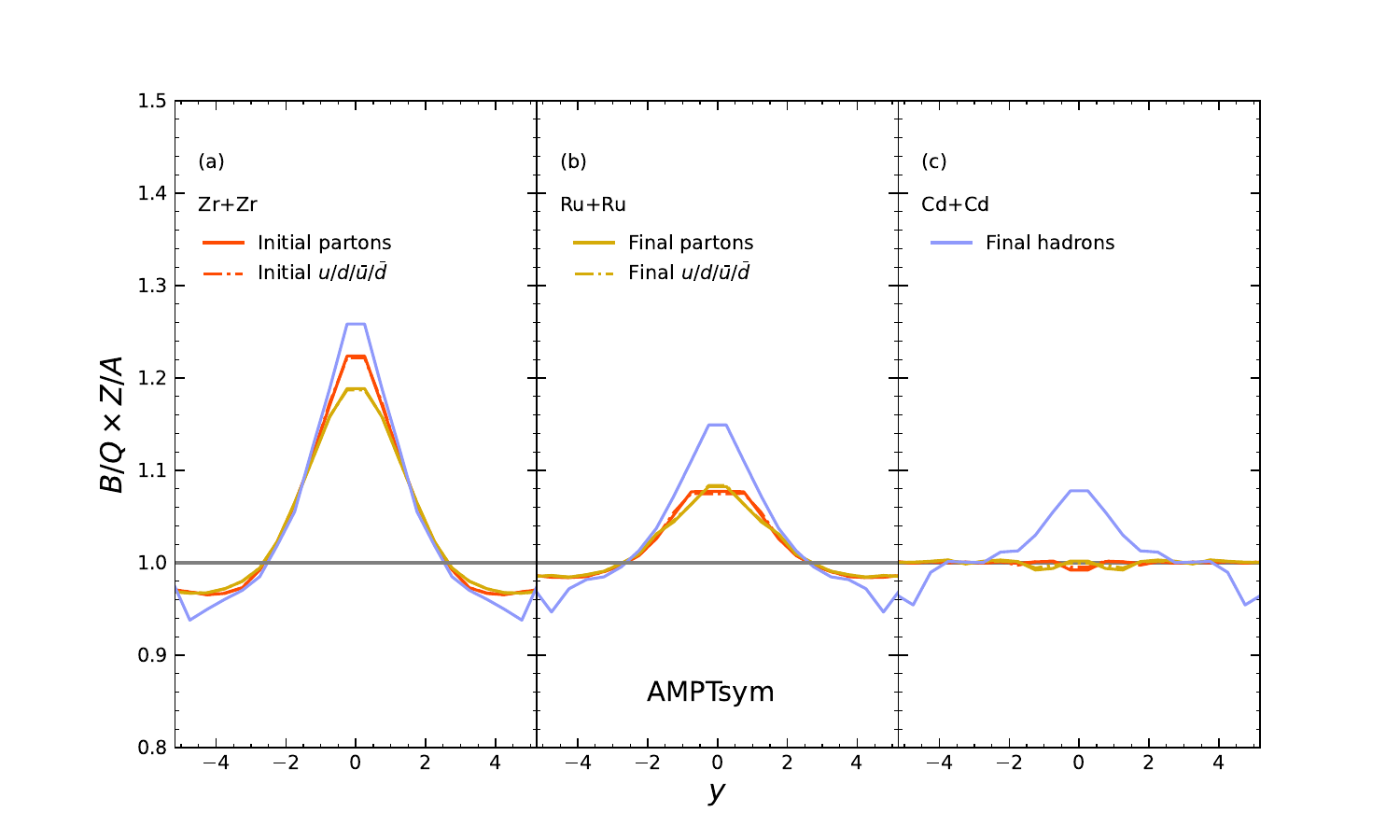}
\caption{Same as Fig.~\ref{fig4} but from a modified AMPT-SM model
  where the distributions of initial strange and anti-strange quarks 
are symmetrized to be the same.}
\label{fig5}
\end{figure}

Since the incoming nucleons carry net $u$ and $d$ quarks, it is
natural for light quarks ($u, d$) to have a different shape of
rapidity distribution  from light antiquarks, as shown in
Fig.~\ref{fig3}. However, it is unclear why and how the rapidity 
distributions of $s$ and $\bar s$ quarks are different from each other; 
naively one may expect their distributions to be the same since $s$
and $\bar s$ are pair produced. To explore the effect of this
uncertainty, we now test a modified version of the AMPT-SM model
(denoted as AMPTsym), where we symmetrize the initial momentum as
well as space-time distributions of $s$ and $\bar s$ quarks before the
parton cascade.  
Results from this model for the $B/Q\times Z/A$ ratios as functions 
of rapidity are shown in Fig.~\ref{fig5} for the three isobar systems. 
For Cd+Cd collisions in panel (c), the initial parton
$B/Q\times Z/A=1.0$ not only for light quarks but also after
including the (anti)strange quarks as expected. 
A big difference from the normal AMPT-SM results is that the AMPTsym
model produces $B/Q\times Z/A>1$ for final hadrons at  mid-rapidity
for all three systems. In addition, we see that the parton cascade has
little effect so that $B/Q\times Z/A \simeq 1.0$ for final partons in
Cd+Cd collisions.  On the other hand, hadronization (quark coalescence 
here) has some effect on the $B/Q\times Z/A$ ratio, e.g., it 
changes the ratio away from unity for Cd+Cd collisions as shown in
Fig.~\ref{fig5}(c). 

\begin{figure}[htb]
\includegraphics[width=0.8\linewidth]{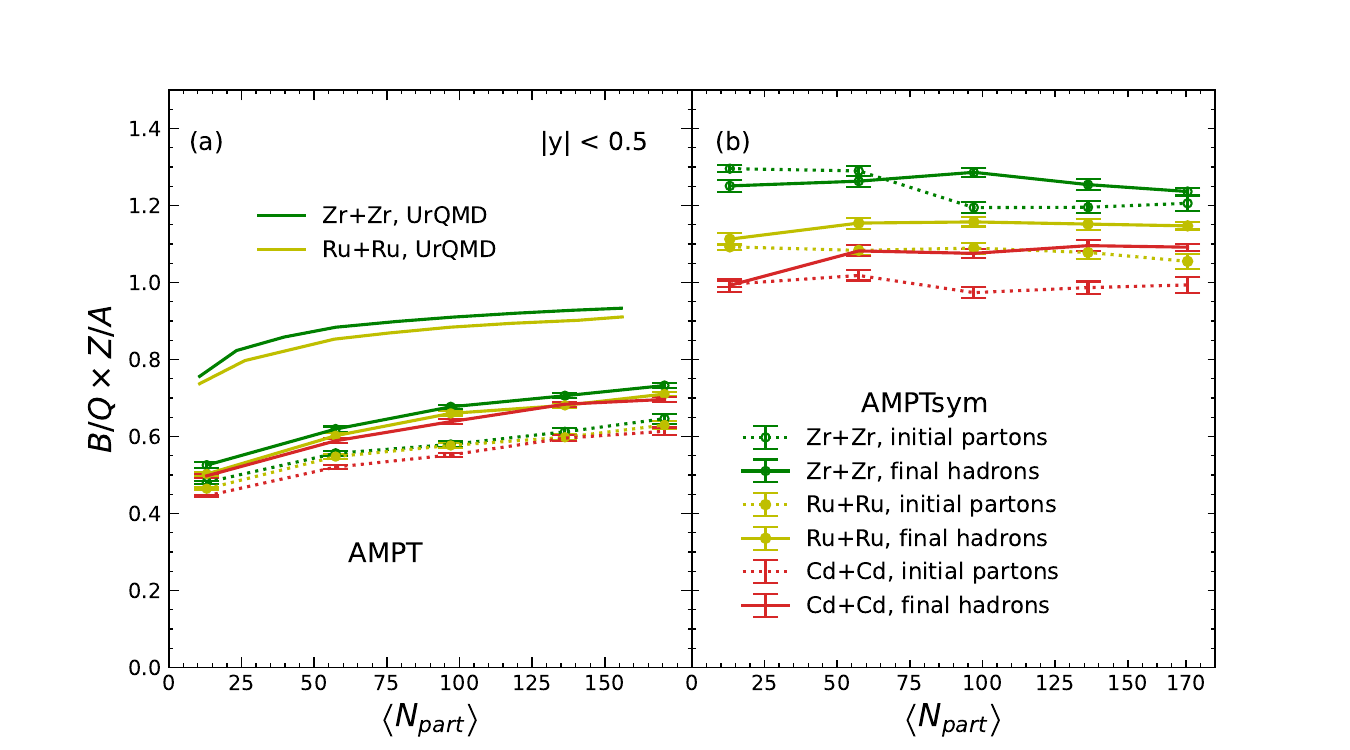}
\caption{The $B/Q\times Z/A$ ratio as a function of $\avgnpart$ for
  isobar collisions from (a) the normal AMPT-SM model and (b) the
  modified AMPT-SM model with symmetrized distributions for initial
  strange and anti-strange quarks. Results for initial partons and
  final hadrons from the AMPT models are shown, and results from the
  UrQMD model are also shown for comparison.}
\label{fig6}
\end{figure}

Figure~\ref{fig6} shows the $B/Q\times Z/A$ ratio at mid-rapidity
as a function of the average number of participant nucleons in isobar
collisions from the AMPT-SM models, where results for initial partons
and final state hadrons are both shown.  
We see  in panel (a) that the ratios from the normal AMPT-SM model 
all gradually increase with $\avgnpart$ but all stay below unity.  
Results from the UrQMD model~\cite{Lewis:2022arg} also show these
features, although they are closer to unity. 
We find that about half of the $B/Q\times Z/A$ increase with
$\avgnpart$ comes  from overall $s-\bar s$ asymmetry as shown in
Fig.~\ref{fig3}, and the other half of the increase with
$\avgnpart$ is due to the modest decrease of the $s-\bar s$ asymmetry
with centrality (where $f_{\bar s}/f_s-1$ decreases from $\sim
9\%$ in peripheral collisions to $\sim 7\%$ in central collisions). 
When we remove the initial $s-\bar s$ asymmetry in the rapidity
distribution, the $B/Q\times Z/A$ ratios at mid-rapidity in panel (b)
are mostly above one, where the ratios for initial partons in Cd+Cd
collisions are consistent with one as expected from Eq.\eqref{bqSym2}.
Therefore, the $s-\bar s$ asymmetry in the initial rapidity
distribution greatly affects the $B/Q\times Z/A$ ratio in nuclear
collisions.

In addition, the $B/Q\times Z/A$ ratio at mid-rapidity depends on the
isospin asymmetry of the incoming nuclei, being the highest
for Zr+Zr collisions and the lowest for Cd+Cd collisions (at the same
$\avgnpart$). This is the case in Fig.~\ref{fig6}  for both initial
partons and final hadrons and for both AMPT-SM models (with or without
the initial $s-\bar s$ asymmetry). This ordering also exists for Zr+Zr
and Ru+Ru collisions from the UrQMD model~\cite{Lewis:2022arg}. 
Since the net-baryon distributions are essentially the same for the
isobar collisions, this ordering results from the net-charge
difference. For example, we find that that the net-electric charge
from light quarks ($u,d, \bar u, \bar d$) in Ru+Ru collisions at
mid-rapidity is about 20$\%$ higher than that from Zr+Zr collisions,
rather than the 10$\%$ we would naively expect from $\Delta Z$.

\subsection{The $B/\Delta Q\times \Delta Z/A$ ratio for isobar 
  systems} 

\begin{figure}[htb]
\includegraphics[width=0.5\linewidth]{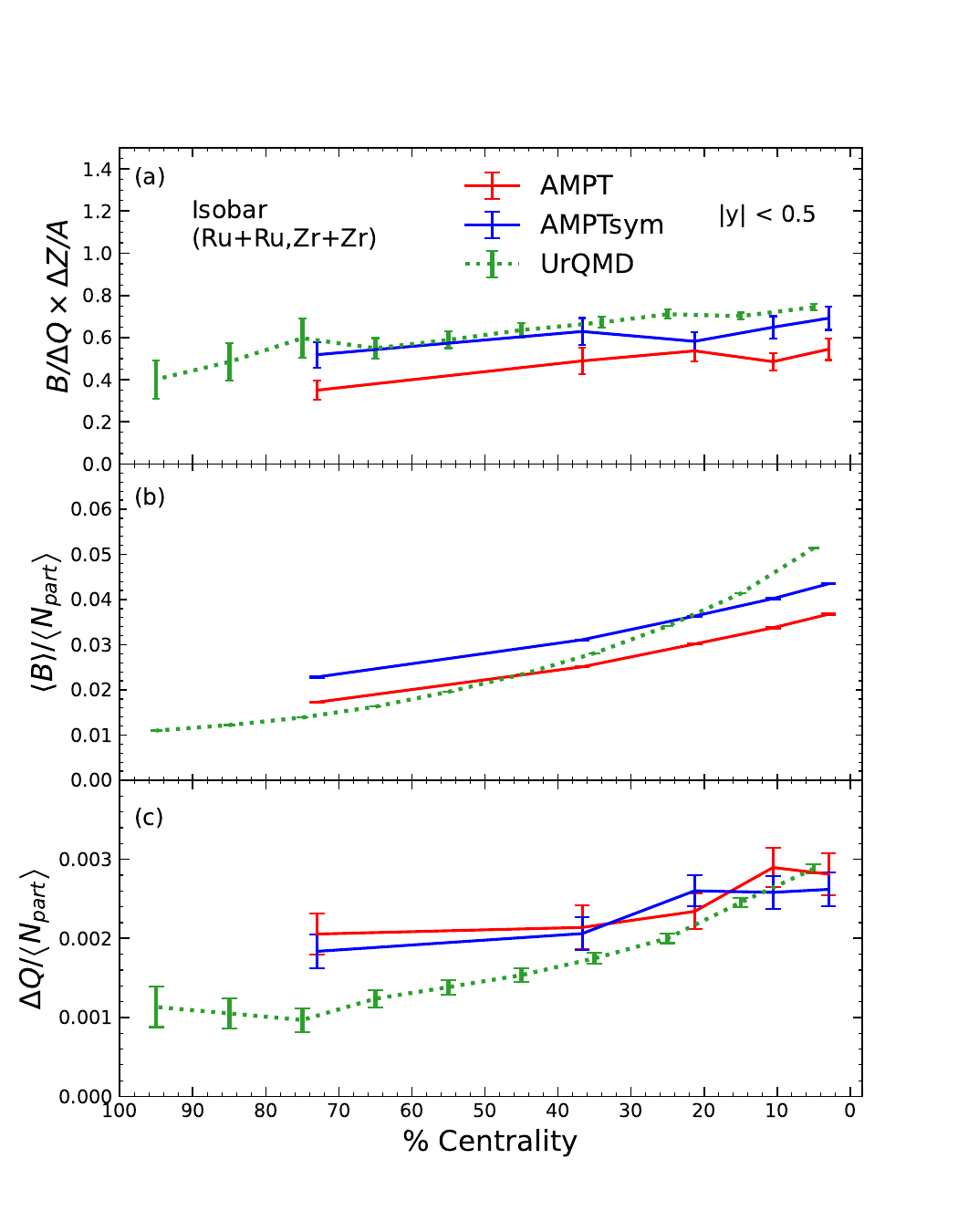} 
\caption{(a) The $B/\Delta Q\times \Delta Z/A$ ratio, 
(b) the average net-baryon number, and (c) the difference of net
electric charge for Ru+Ru and Zr+Zr collisions at mid-rapidity (scaled
by $\avgnpart$) from the normal and $s-\bar s$ symmetrized AMPT-SM
models versus centrality; UrQMD   results~\cite{lv:2025} are also
shown for comparison.}
\label{fig7}
\end{figure}

We now examine the difference between Ru+Ru and Zr+Zr isobar
collisions in Fig.~\ref{fig7}(a) by plotting the $B/\Delta Q\times
\Delta Z/A$ ratio of final hadrons at mid-rapidity as a function of 
centrality.  We see that the ratios from both the normal
AMPT-SM model and the AMPTsym model are well below unity at
mid-rapidity, where the ratios from the AMPTsym model are higher 
and close to those from the UrQMD model. 
On the other hand, a recent study by the STAR
Collaboration~\cite{STAR:2024lvy} showed 
that the $B/\Delta Q\times \Delta Z/A$ data for isobar collisions are
much higher (mostly by a factor of 2 or more) than all the model
results shown in Fig.~\ref{fig7}(a) for all centralities. 
Therefore, the puzzle about the baryon versus charge stoppings
in Ru+Ru and Zr+Zr isobar collisions still exists, regardless of the
uncertainty about the initial $s-\bar s$ asymmetry. 

We also show the average net-baryon
number in Fig.~\ref{fig7}(b) and the net-electric charge difference
($\Delta Q \equiv Q_{\rm Ru+Ru}-Q_{\rm Zr+Zr}$) in Fig.~\ref{fig7}(c)
at mid-rapidity for isobar collisions as functions of centrality, with
both  being scaled by $\avgnpart$. We see that the AMPT results are
different from the UrQMD results while they overlap at certain centralities. 
The $\avgb/\avgnpart$ data values from the STAR
Collaboration~\cite{STAR:2024lvy} are significantly higher than the
model  results in peripheral collisions while they are close to the
values from the AMPTsym  and UrQMD models for central
collisions. On the other hand,  the STAR $\Delta Q/\avgnpart$ data 
are close to the  model results in peripheral collisions but are much
lower (by about a factor of 2) than the model results for central
collisions. As a result, the $B/\Delta Q\times \Delta Z/A$ STAR data
at mid-rapidity  are much higher than all the model results shown in 
Fig.~\ref{fig7}(a) at all centralities.  Note that the UrQMD results
shown in Fig.~\ref{fig7} use the UrQMD $\avgnpart$
values~\cite{lv:2025} and are thus not the same as the UrQMD curves
shown in the recent STAR study~\cite{STAR:2024lvy}.  

\begin{figure}[htb]
\includegraphics[width=0.5\linewidth]{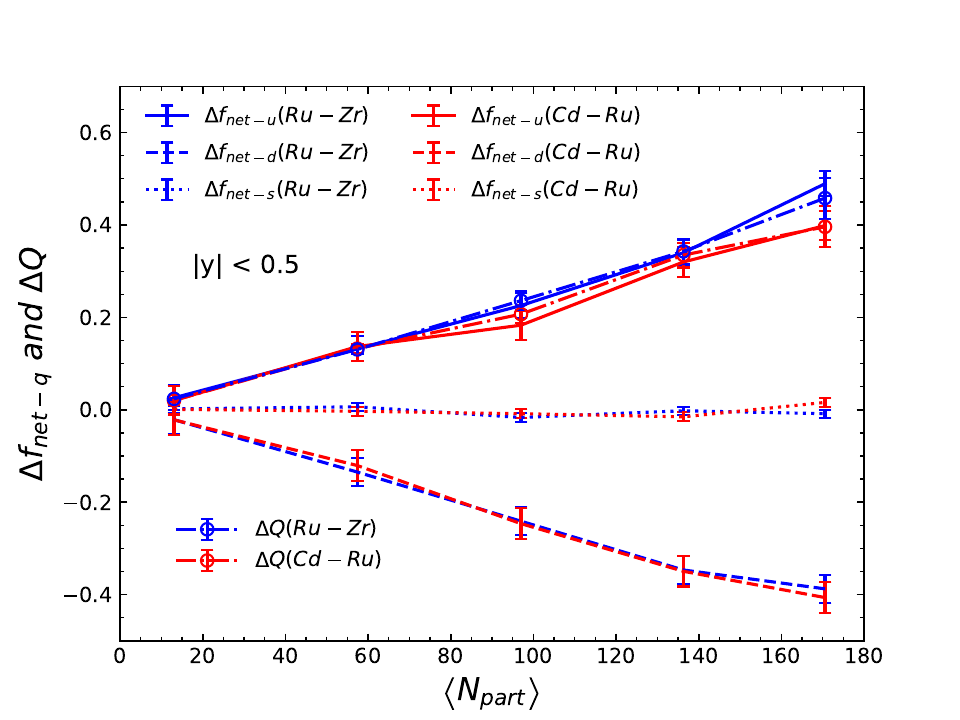}
\caption{The differences of the net-quark number of different flavors
  and the net-electric charge at mid-rapidity from the AMPT-SM model
  as   functions of $\avgnpart$  between Ru+Ru and Zr+Zr isobar
  collisions   and between Cd+Cd and Ru+Ru isobar collisions.} 
\label{fig8}
\end{figure}

When the isobar collision system changes from Zr+Zr to Ru+Ru, we can
write the changes of the net-baryon number and the net-electric charge
at a given rapidity as
\begin{eqnarray}
\Delta B=(\Delta f_{net-u}+\Delta f_{net-d}+\Delta f_{net-s})/3,
  ~\Delta Q=(2\Delta f_{net-u}-\Delta f_{net-d}-\Delta f_{net-s})/3, 
\end{eqnarray}
where $\Delta f_{net-u} = \Delta f_u -\Delta f_{\bar u}$. 
As shown in Fig.~\ref{fig8}, we find from the normal AMPT-SM model
that $\Delta f_{net-s} \simeq 0$ between Ru+Ru and Zr+Zr collisions
(and also between Cd+Cd and Ru+Ru collisions). Given the observation 
$\Delta B \simeq 0$ among the three isobar systems, we then have
$\Delta f_{net-u} \simeq -\Delta f_{net-d}$ and $\Delta Q \simeq
\Delta f_{net-u}$,  both of which are confirmed by Fig.~\ref{fig8}. 
This is why the normal AMPT model and the AMPTsym model give about  
the same $\Delta Q/\avgnpart$ results in Fig.~\ref{fig7}(c). 
At the parton level, we then get  
\begin{eqnarray} 
B/\Delta Q\times \Delta Z/A 
\simeq \frac{\Delta Z}{A} \frac{B}{\Delta f_{net-u}} 
\simeq \frac{\Delta Z}{A}\left (\frac{2f_{net-q}}{\Delta f_{net-u}}
  \right ) \left ( 1+\frac{f_{net-s}}{2f_{net-q}} \right ).
\label{bdq}
\end{eqnarray}
Since $\Delta Q$ essentially does not depend on the (anti)strange 
distributions, the isobar $B/\Delta Q\times \Delta Z/A$ ratio is 
very sensitive to the net-light quark ($u,d$) 
stoppings and their change between the two isobar systems,  
but it is less sensitive to (anti)strange distributions and the $s-\bar
s$ asymmetry. This is very different from the $B/Q\times Z/A$ ratio,
which is very sensitive to the $s-\bar s$ asymmetry. 
Comparing the above Eq.\eqref{bdq} with Eq.\eqref{bq}, we find that
the $B/Q\times Z/A$ ratio is more sensitive to the  
$s-\bar s$ asymmetry by a factor of 3 than the 
$B/\Delta Q\times \Delta Z/A$ ratio. 

\section{Conclusion}

The stopping of baryons in nuclear collisions has not been well
understood, and it is closely related to the question of whether a quark
or a gluon junction carries the baryon number. It is also possible
that (anti)quarks carry the baryon number in general, while a gluon
junction that joins three quarks together topologically
carries the baryon number of a color-singlet baryon in the confined
phase and thus affects the baryon number transport in nuclear
collisions.  Recently it has been proposed that comparing the
net-baryon with the net-electric charge in nuclear collisions can help 
address the question, e.g., with the $B/Q\times Z/A$ ratio at
mid-rapidity. Here we study the baryon and charge stoppings in nuclear
collisions, where it is known that the $B/Q\times Z/A$ ratio can
strongly depend on the rapidity, although its value is one for the
full phase space. We find that the $B/Q\times Z/A$ ratio depends
crucially on the difference between strange and anti-strange quark
rapidity distributions (the $s-\bar s$ asymmetry), and having several
percent more anti-strange than strange quarks at mid-rapidity would
lead to a ratio well below one for Zr+Zr and Ru+Ru isobar collisions.  
This is why the string melting version of the AMPT model
gives $B/Q\times Z/A <1$ at mid-rapidity. On the other hand, if the
initial strange and   anti-strange quarks have the same rapidity 
distribution, the $B/Q\times Z/A$ ratio at mid-rapidity becomes one
or higher for isobar collisions.  For collisions of isospin symmetric
nuclei such as $^{40}\rm Ca$, the $B/Q\times Z/A$
ratio would deviate from unity only because of this $s-\bar s$
asymmetry, when the neutron skin effect is neglected. 

In addition, we find that the $B/\Delta Q\times \Delta Z/A$ ratio
depends crucially on the net-light quark ($u,d$) stoppings and their
change between the two isobar systems,  but it is less sensitive to
the $s-\bar s$ asymmetry than the $B/Q\times
Z/A$ ratio by a factor of 3. 
For isobar collisions, the AMPT model gives $B/\Delta Q\times \Delta
Z/A <1$, regardless of whether the $s-\bar s$ asymmetry is
removed. Since a recent STAR Collaboration study shows $B/\Delta
Q\times \Delta  Z/A$ data much higher than the values from the AMPT
model and the UrQMD model, there is an interesting puzzle about baryon
versus  electric charge stoppings.  Experimental data on the net-baryon
and net-electric charge from different hadron species including
strange and non-strange hadrons would  provide more information to
constrain the unmeasured $B/Q\times Z/A$ ratio and advance our
understanding of baryon and charge stoppings.

\section*{Acknowledgments}
We thank Drs. Rongrong Ma, Zebo Tang and Zhangbu Xu for helpful 
discussions. We also thank Wendi Lv for providing the UrQMD
results. This work has been supported by the National Science
Foundation under Grant No. 2310021.

\bibliography{refs}

\end{document}